\documentclass[3p,times]{elsarticle}

\makeatletter
\def\ps@pprintTitle{%
 \let\@oddhead\@empty
 \let\@evenhead\@empty
 \def\@oddfoot{}%
 \let\@evenfoot\@oddfoot}
\makeatother

\usepackage{graphicx}
\usepackage{amssymb}
\usepackage{amsthm}
\usepackage{amsmath}
\usepackage{float}
\usepackage{subcaption}

\usepackage{verbatim}

\usepackage{xcolor}

\usepackage{lineno}

\biboptions{comma,round,authoryear}

\journal{tbd}

\begin{document}

\begin{frontmatter}


\title{Vertical vs. Horizontal Policy in a Capabilities Model of Economic Development}


\author{Alje van Dam \fnref{label1,label2}}
\ead{A.VanDam@uu.nl}

\author{Koen Frenken \fnref{label1}\corref{cor1}}
\ead{K.Frenken@uu.nl}


\fntext[label1]{Copernicus Institute of Sustainable Development, Utrecht University}

\fntext[label2]{Center for Complex Systems Studies, Utrecht University}

\cortext[cor1]{Corresponding author}



\begin{abstract}
Against the background of renewed interest in vertical support policies targeting specific industries or technologies, we investigate the effects of vertical vs. horizontal policies in a combinatorial model of economic development. In the framework we propose, an economy develops by acquiring new capabilities allowing for the production of an ever greater variety of products with an increasing complexity. Innovation policy can aim to expand the number of capabilities (vertical policy) or the ability to combine capabilities (horizontal policy). The model shows that for low-income countries, the two policies are complementary. For high-income countries that are specialised in the most complex products, focusing on horizontal policy only yields the highest returns. We reflect on the model results in the light of the contemporary debate on vertical policy.

\end{abstract}


\end{frontmatter}

\newpage

\section{Introduction}
\label{S:1}
Vertical policy is back on policy agendas globally \citep{Rodrik2004, Cimoli2009, Chang2020}. Regarding low-income countries, vertical policies comprise a variety of instruments to allow a country to catch up with the global technology frontier. One popular strategy historically has been to temporarily protect 'infant industries' from global competition as to build up knowledge capabilities and institutions required for developing particular technologies or industries \citep{Freeman1987, Chang2002}. A more recent approach has become known as modern industrial policy, and attempts to back entrepreneurs who discover new export industries with complementary public investments \citep{Rodrik2004}. Evaluation studies confirmed the positive role that vertical policies policies can play in fostering economic development, though success depends on the exact policy design and contextual conditions \citep{Lane2020}. These insights, combined with the spectacular success of China fully embracing vertical policy, has put vertical policy high on the policy agenda as a strategy for economic development for low-income countries.

Vertical policy is also experiencing popularity in high-income countries \citep{Aiginger2007, Aghion2011}. In the light of disappointing growth rates, the effectiveness of horizontal policies is increasingly questioned \citep{Mazzucato2011, Mazzucato2015}. In high-income contexts, vertical policy is called for to push the technological frontier itself rather than to catch-up with technologies already developed in other countries. Vertical policies come in different versions and with different labels, including industrial policy, policies for key enabling technologies, smart specialisation, transformative innovation policy, and mission-oriented innovation policy. Though the rationales and instruments tend to differ for each of these policies, but they share a vertical orientation towards supporting only specific industries or technologies \citep{Foray2019, Mazzucato2018, Bailey2019}.

The renewed interest in vertical innovation policy and the proliferation of new policy concepts has not been matched with new theoretical frameworks. The lack of theorising is in itself not surprising given that innovation and development are complex and elusive phenomena. What is more, economic growth models have long neglected the role of the exact industries or technologies in an economy, and the process of diversification leading to new industries and technologies. However, with the recent advent of a new capability theory of economic growth as developed by Hausmann and others \citep{Hausmann2007, Hidalgo2009, Hausmann2011, Inoua2016, Sutton2016, VanDam2020}, a new framework has become available to theorise about policy and its effects on economic development. This paper sets out to develop a policy framework based on the capability theory of economic growth as to assess and compare the returns of vertical and horizontal policies. 

The capability theory starts from an explicit representation of specific outputs and the inputs required to produce each output. Outputs are generally considered (export) products and inputs as 'capabilities', which include assets, knowledge and skills, but also products-specific regulations and institutions \citep{Lall2000, Hidalgo2009, Hausmann2011}. Economic development stems from diversification into new products made possible by the acquisition of new capabilities. Once acquired, firms start recombining the new capability with existing ones, thus increasing both the variety of products (number of products) and complexity of products (number of capabilities used in each product) in the economy. A vertical policy can be thought of as any policy that targets the acquisition of a particular capability. For example, an industrial policy focusing on aircraft production, would lead to the acquisition of one or more new capabilities, which - combined with already existing capabilities - enable a country to start producing aircraft. Once a new capability is acquired, it can also be used in other recombinations of inputs allowing further diversification into new products. The practical challenge for any vertical policy, then, is to target a capability that can be effectively recombined with existing capabilities as to increase the variety and complexity of an economy (e.g., industrial policy. targeted R\&D investment, new teaching programs, selective Foreign Direct Investment, selective migration policy, etc.).

The capability theory of economic development is, in its current form, still a limited framework as it stands on two strong assumptions. First, it assumes that that countries produce every product that their capabilities base would enable them to produce. This assumption is at odds with the common observation that high-income countries lose industries to countries with lower wages over the product lifecycle \citep{Vernon1966}. If one instead assumes that countries stop producing low-complexity products as the average complexity of their products continues to increase with the acquisition of new capabilities, it can be shown that, over time, the trend of increasing variety changes into a trend of decreasing variety, consistent with the empirical phenomenon of the hump \citep{Cadot2011, Sutton2016, VanDam2020}.

The second strong assumption in capability models is that countries would not face any limitation in being able to recombine capabilities. Put differently, it views countries as having unlimited abilities to effectively coordinate any number of capabilities required for a product. It follows from this assumption that the only objective for a policy maker would be to acquire new capabilities. If so, the policy question boils down to selecting which capabilities should be acquired and in what manner. Once one would relax this assumption and would view countries as facing constraints in the complexity of products that their firms are able to make, a more fundamental policy question arises: how much effort should a country put on acquiring a specific new capability vs. how much effort should it put on learning how to make more complex products from the capabilities already present. It is the latter policy that we will consider as a horizontal policy, which aims to increase the ability of a country to produce more complex products. Here, horizontal policy refers all policies that improve the coordination and integration of capabilities required for the production of products (e.g., basic research, public research organizations, standardization institutes, public consultation schemes, collaboration subsidies, generic social and managerial skills, laws, and regulations), similar to what has been referred to as a country's 'national innovation system' \citep{Freeman1987, Lundvall1992}. 

It follows that policy for economic development can be understood as a combination of two policies: a vertical policy focusing on acquiring a new capability providing a country with opportunities to produce a larger variety of products, and a horizontal policy focusing on improving a country's ability to recombine capabilities in ever more complex products. Given these two types of policies, the question then becomes how to allocate their efforts on one or the other policy. Intuitively, one may expect the two policies to be complementary: the combinatorial logic of products stemming from combinations of capabilities implies that the ability to recombine capabilities is most valuable for countries that already have many capabilities.

Building on previous combinatorial models of economic development \citep{Hausmann2011, Inoua2016, VanDam2020}, we propose a model in which we conceive of economic development as the outcome of increases in the number of capabilities residing in a country and of improvements in the ability to recombine capabilities in a country. National government decides, at each time step, whether to increase the number of capabilities in a country or to improve the ability to recombine capabilities. This decision depends on the expected increase in the average complexity of products. This basic model set-up will explain the complementarity between vertical and horizontal policy. We then turn to our extended model by introducing a minimum wage that bounds the minimum complexity of products produced in a country. As a country enters more complex products, it increases its minimum wage and abandons its products with lowest complexity. This extension of the model leads to three further contributions. First, the resulting model reproduces the stylised fact of the hump. Second, it explains the growing importance of horizontal policies over vertical policies as economies develop over time. Third, it can localise the shift in optimal policy close to the hump, suggesting that high-income countries should focus on horizontal rather than vertical policies.

\section{The Model}
\label{S:2} 
Our understanding of economic development has long been guided by the notion of a production function that specifies how inputs such as capital and labor translate into the total output of an economy. More recently, models are more explicit about the products produced in an economy. At the level of products, inputs can be considered to be strictly complementary \citep{Kremer1993, Hausmann2011, Brummitt2017}. This assumption is based on the idea that the production of any product or service requires a particular combination of complementary inputs. 

Inputs required to produce a product have been referred to as 'capabilities' \citep{Hidalgo2009, Hausmann2011}. Following this reasoning, the ability of an economy to produce a product depends on the capabilities present in a country. Developing new products consists of recombining old and new inputs into configurations that have economic value \citep{Inoua2016}. It also follows that with the acquisition of a new capability, the variety of products that a country can produce grows in a non-linear fashion. An elementary model of this kind is that each possible combination of capabilities results in one unique product. The total number of products that a country can make is then given by summing the number of possible combinations of a given length that can be made out of $n$ capabiltiies over all possible lengths $s$:
\begin{align*} 
d(n) = \sum_{s=0}^n {n \choose s} = 2^n. 
\end{align*} 
The average complexity of products is given by the total length of all products divided by the total number of products:
\begin{align*} 
\bar{s}(n) = \frac{\sum_{s=0}^n s {n \choose s}}{2^n} = \frac{n}{2}.
\end{align*} 

The assumption that any combination of capabilities leads to a viable product is arguably too strong. Instead, one can safely assume that only some combinations of capabilities lead to meaningful products. The set of combinations of capabilities resulting in meaningful products has been referred to as a 'recipe book', which describes the combinations of capabilities that are complementary in that they lead to viable products \citep{Hausmann2011, Inoua2016, Fink2017}.

The model can be generalized by assuming that every capability is part of a viable product with a given probability $\rho$ \citep{Inoua2016, VanDam2020}. A combination of $s$ capabilities then has probability $\rho^s$ of representing a viable product of length $s$. Hence, it becomes increasingly unlikely that a combination of capabilities leads to a viable product as more capabilities are added, since $\rho^s$ is decreasing in $s$ when $\rho<1$. For $\rho=1$, we recover the initial simple model described above.

Since there are ${n \choose s}$ possible combinations of $s$ components one can make from the total of $n$ components, and each combination of length $s$ has probability $\rho^s$ of being viable, the expected number of products of length $s$ a country with $n$ components can make is given by $d(n,s) = {n \choose s} \rho^s$. Summing this quantity over all product lengths $s$ gives the expected variety of products that can be made with $n$ components
\begin{align*}
d(n) = \sum_{s=0}^n {n \choose s} \rho^s = (1+\rho)^n.
\end{align*} 
Since the share of products of length $s$ in a country is given by $\frac{{n \choose s} \rho^s }{d(n)}$, the expected average complexity given $n$ components can be computed as \citep{Inoua2016, VanDam2020}
\begin{align} \label{eq:avgs}
\bar{s}(n) =  \sum_{s=0}^n s \frac{{n \choose s} \rho^s}{d(n)} &=\frac{\rho}{1+\rho}n.
\end{align}

Note that while variety increases exponentially with $n$, complexity increases only linearly with $n$. The rate of increase in product complexity viz. economic growth is solely determined by the difficulty parameter $\rho$.

\section{Vertical vs. horizontal policy}
\label{S:3}

The key assumption in the combinatorial model, albeit an implicit one, holds that a country can recombine any number of capabilities. That is, the sole challenge for a country is to acquire additional capabilities, leading to an increase in $n$, which automatically translates into a stable growth path in the form of a linear increase in average product complexity.

\begin{figure}[h]
\centering 
\includegraphics[width=\textwidth]{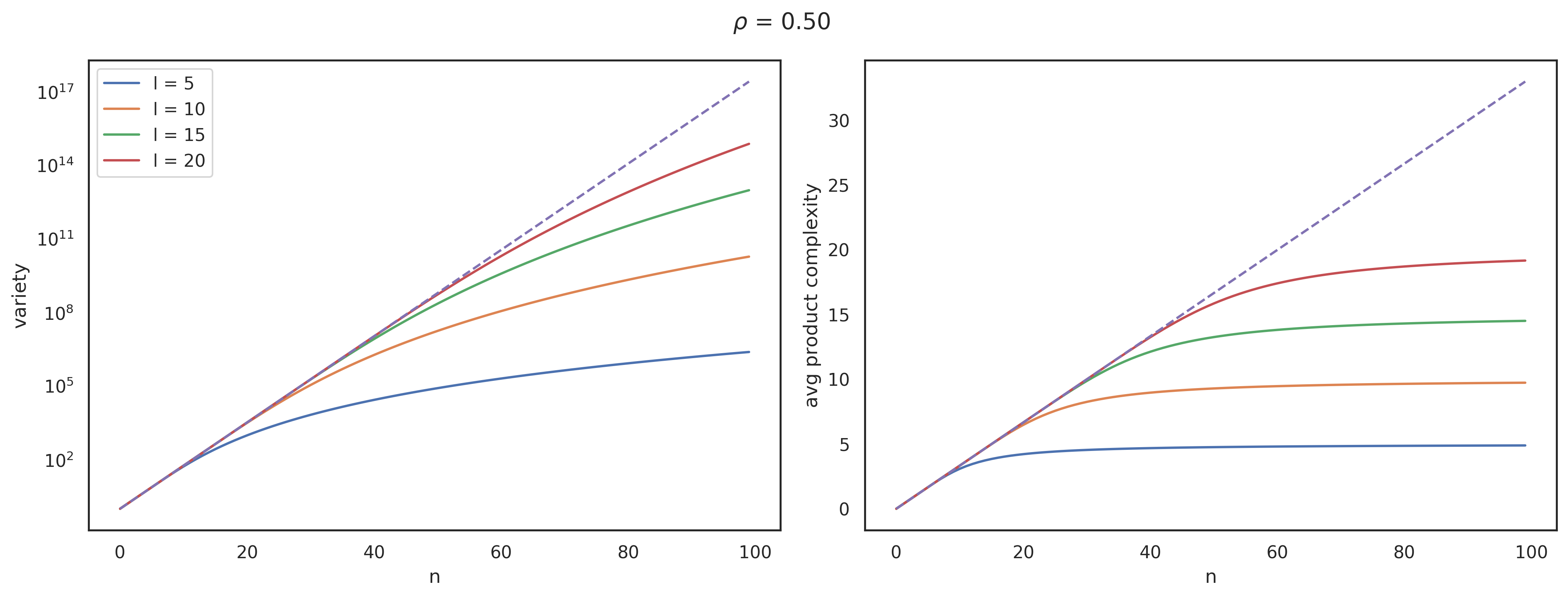}
\caption{Limits to coordination} 
\label{fig:limits} 
\end{figure}

Dropping the assumption that countries can recombine any number of capabilities, we introduce the parameter $l$ referring to the maximum length of products that a country is able to produce. The expected product variety and product complexity are then given by
\begin{align*}
    d(n,l) &= \sum_{s=0}^l {n \choose s} \rho^s \\
    \bar{s}(n,l) &= \frac{\sum_{s=0}^l s {n \choose s} \rho^s}{d(n,l)},
\end{align*}
respectively. Figure \ref{fig:limits} shows how a constraint on the maximum complexity of products, as expressed by $l$, hampers economic development as product variety (left) grows less than exponentially and average product complexity (right) reaches a ceiling converging asymptotically to $l$ with $n$ approaching infinity.  

A policy maker now has two options to foster economic development. First, (s)he can increase the number of capabilities $n$, to which we refer as vertical policy. We model the decision to increase the number of capabilities as a unit increase in $n$. The second option is to improve the country's ability to recombine capabilities, to which we refer to as horizontal policy. We model a decision to increase the ability to recombine capabilities as a unit increase in $l$.

Note at this point that our model remains agnostic about the specific type of vertical policy that is being employed. Rather, we model vertical policy as any policy that leads to some new capability that has random complementarities with already existing capabilities. In this sense, vertical policies are targeted only in the sense that they lead to one new capability, but blind with regard to the exact complementarities that can be exploited between the new capability and the already existing ones. 

Starting with the initial condition in a country with $n$=1 and $l$=1, the policy maker alternates between the two policies depending on which policy yields the highest expected increase in average complexity. For vertical policy, the expected gain in average product complexity from adding a capability is given by 
\begin{align*}
    \frac{\Delta \bar{s}}{\Delta n} &= \bar{s}(n+1,l) - \bar{s}(n,l) \\
    &= \frac{\sum_{s=0}^l s {n+1 \choose s} \rho^s}{d(n+1,l)} - \frac{\sum_{s=0}^l s {n \choose s} \rho^s}{d(n,l)}.
\end{align*}
For a horizontal policy, i.e. increasing $l$, the gain in average product complexity is given by
\begin{align*}
     \frac{\Delta \bar{s}}{\Delta l} &= \bar{s}(n,l+1) - \bar{s}(n,l) \\
     &= \frac{\sum_{s=0}^{l+1} s {n \choose s} \rho^s}{d(n,l+1)} - \frac{\sum_{s=0}^l s {n \choose s} \rho^s}{d(n,l)}.
\end{align*}
At any given stage in the development process (characterized by $n$ and $l$), a policymaker chooses for vertical policy when $\frac{\Delta \bar{s}}{\Delta n} >  \frac{\Delta \bar{s}}{\Delta l}$, and for horizontal policy when $\frac{\Delta \bar{s}}{\Delta l} > \frac{\Delta \bar{s}}{\Delta n}$. 




Following this policy decision model, we simulated the evolution of product variety and average product complexity over time (upper left and middle left panel in Figure \ref{fig:panel_0.50}) as well as the incidence rates of vertical policy and horizontal policy (lower left panel in Figure \ref{fig:panel_0.50}). The two policies are clearly complementary as vertical policies (increasing $n$) are alternated by horizontal policies (increasing $l$) as to leverage the increased potential to make more complex products due to the recent rise in capabilities. We further observe that the exact incidence rates of both policies are sensitive to $\rho$ (compare lower left panel of Figures \ref{fig:panel_0.50}, \ref{fig:panel_0.25} and \ref{fig:panel_0.75}).

\section{Full model}
\label{S:4}

While our model explains the complementarity between vertical policy and horizontal policy, it falls short in reproducing the the inverted-U shape relationship between income per capita and product variety commonly known as 'the hump'. In terms of economic development, this pattern indicates that countries first diversify and then, at some level of income, start specialising again \citep{Imbs2003, Cadot2011}.

In our combinatorial framework, the hump can be understood as resulting from low-complexity products exiting a country's portfolio as a country continues to diversify into high-complexity products \citep{VanDam2020}. Labour involved in low-complexity products arguably has lower productivity, resulting in lower wages, than labour involved in high-complexity products. Economic development leading to products with higher complexity will thus push the highest wages in a country upwards. Assuming minimum wages to increase with maximum wages, a country cannot remain competitive in low-complexity products and will lose these product to low-wage countries.

Implementing such a mechanism of product exit in our model, we assume that countries only produce products with a complexity in the range of [$l$ - $r$, $l$], where $r>0$. This range is based on the idea that given the minimum and maximum wage in a country, it can only be competitive in a certain range of product complexities. It follows that once $l>r$, a country starts abandoning products with the lowest complexity from its portfolio.

\begin{figure}[h]
\centering 
\includegraphics[width=\textwidth]{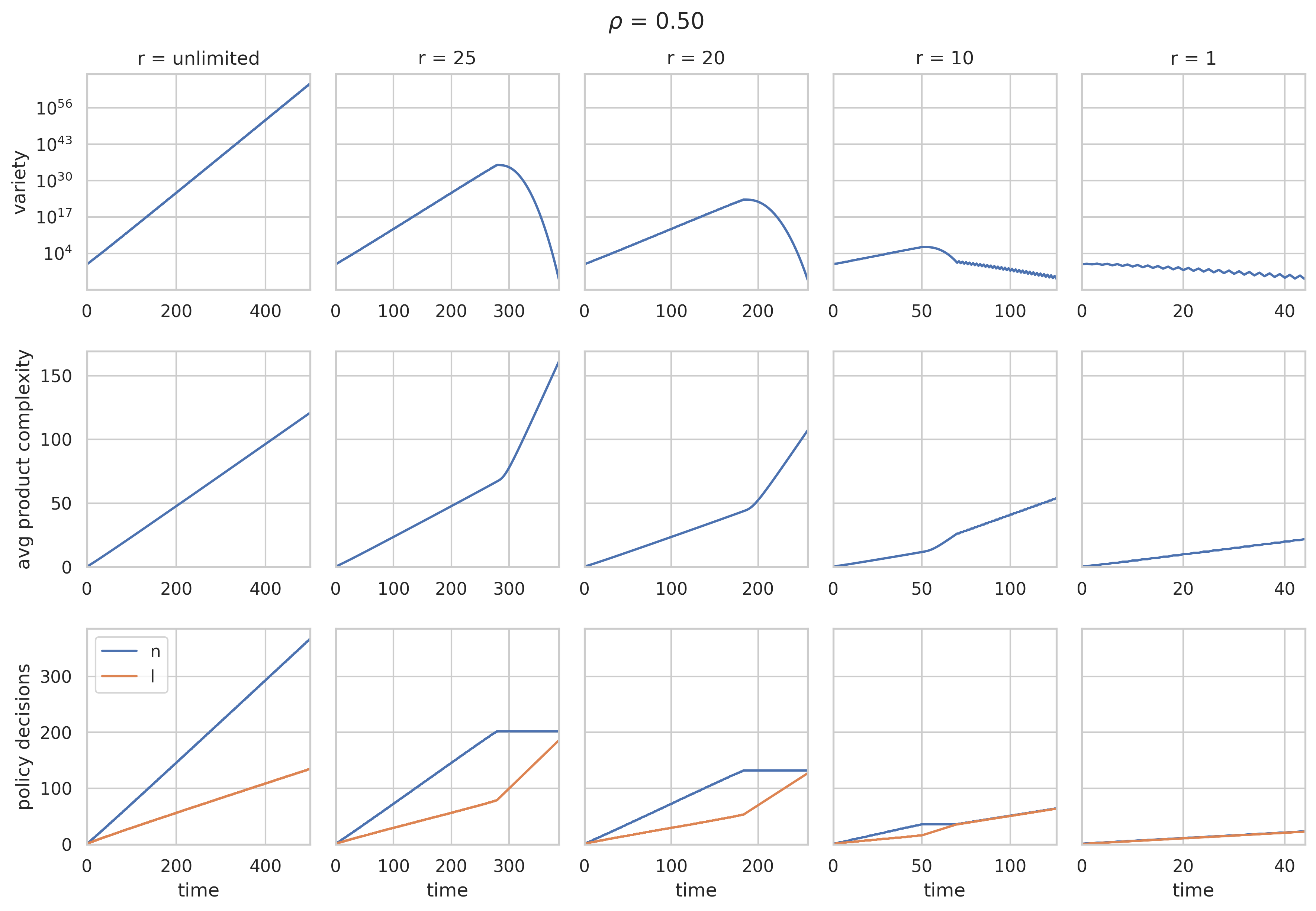}
\caption{Model results for $\rho = .5$.} 
\label{fig:panel_0.50} 
\end{figure}

\begin{figure}[h]
\centering 
\includegraphics[width=\textwidth]{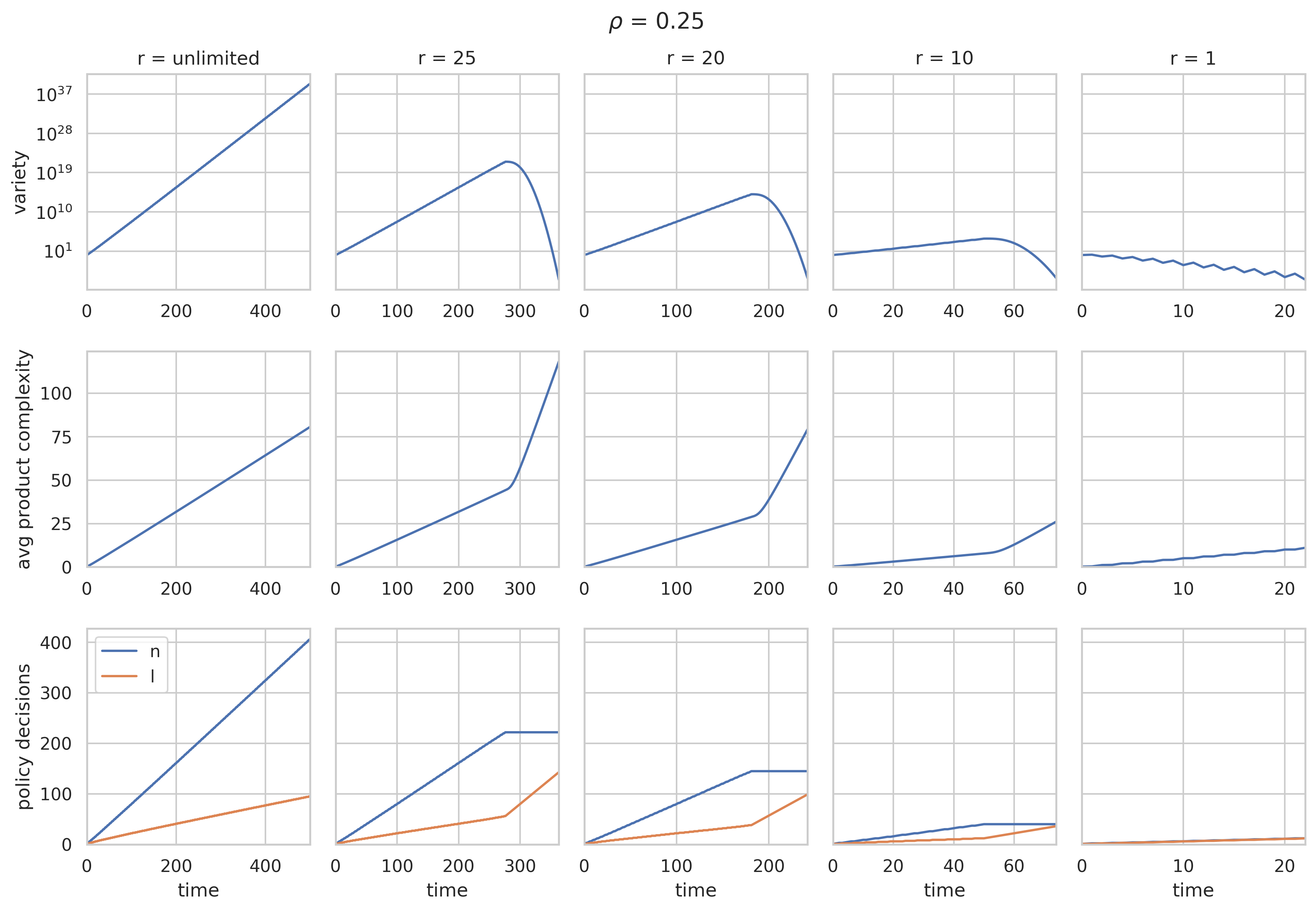}
\caption{Model results for $\rho = .25$.} 
\label{fig:panel_0.25} 
\end{figure}

\begin{figure}[h]
\centering 
\includegraphics[width=\textwidth]{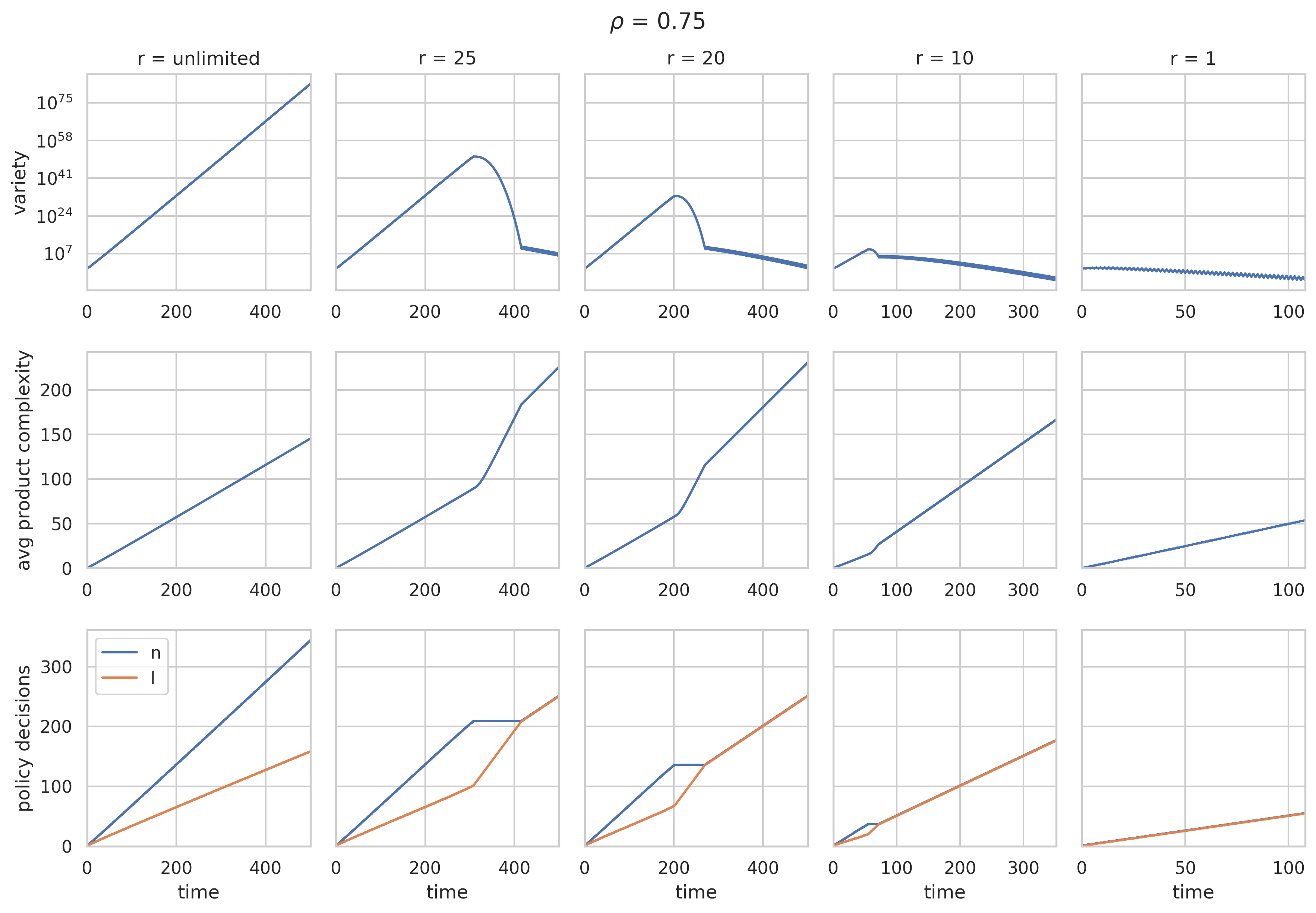}
\caption{Model results for $\rho = ..75$.} 
\label{fig:panel_0.75} 
\end{figure}

The second to fifth columns in Figure \ref{fig:panel_0.50} show the results when we re-run the baseline model (shown in the first column), but now including parameter $r=25$, $r=20$, $r=10$ and $r=1$ respectively. Figures \ref{fig:panel_0.25} and \ref{fig:panel_0.75} show the same results, but now for $\rho=0.25$ and $\rho = 0.75$.

Three observations can be made. First, the model reproduces the hump for non-trivial values of $r$ ($r=25$, $r=20$, $r=10$), as can be seen in the first row of each figure. This is consistent with the empirical phenomenon of the 'hump' \citep{Imbs2003, Cadot2011}, and reproduces the theoretical result of a similar capability model by \citet{VanDam2020}.


The second observation to be made is that product complexity starts accelerating once the variety of a country starts decreasing. Once $l>r$, a horizontal policy will then push complexity upwards in two ways: the policy increases the maximum product complexity that a country is able to produce ($l$) by one and it increases the minimum product complexity that is being produced in a country ($l-r$) by one.

The final observation to be made holds that during the 'hump-period', the optimal policy is to focus solely on horizontal policies, which maximizes the increase in product complexity. Such policy leverages the high number of capabilities already present by improving a country's ability to recombine its capabilities in ever more complex products. This process continues until $l=n$, reflecting a most 'advanced' economy producing solely the most complex products within the range [$n-r$, $n$]. It is also at this stage that vertical policy becomes relevant again next to horizontal policy, as further progress can only be reached by alternating between adding a capability and increasing maximum complexity. Importantly, the focus on horizontal policy in the hump-period is robust for different values of parameters $\rho$ and $r$. And, as the hump phenomenon historically tends to occur only at a certain levels of income per capita, our model can pinpoint the countries that, on theoretical grounds, could benefit most from focusing on horizontal policies (the hump tends to occur at around 24,000 US Dollar (PPP in constant 2000) \citep{Cadot2013}).

\section{Discussion}
\label{S:5}

Elaborating on the capabilities framework of economic development proposed by \citet{Hausmann2011}, \citet{Inoua2016} and \citet{VanDam2020}, we have modelled an economy as developing over time by acquiring new capabilities one-by-one. Every new capability can, with some probability, be recombined with existing capabilities to allow for the production of an ever greater product variety and product complexity. Different from previous models, however, we pose that countries are potentially constrained in the level of product complexity they can handle, due to an under-investment in basic research, managerial skills and an underdeveloped 'innovation system'. 

It follows from our model that public policy can focus on two development strategies: the addition of a new capability, which we refer to as vertical policy, or an improvement of a country's generic ability to recombine capabilities, which we refer to as horizontal policy. The key result that we draw from the model is that for low-income countries, vertical policy focused on capability acquisition is to be complemented with horizontal policy so the increasing number of capabilities can be effectively recombined in more valuable products. A second insight holds that once a country starts abandoning low-complexity products from its portfolio, horizontal policy becomes even more important. In this stage, a country loses competitiveness in relatively simple products, and needs to focus on mastering the coordination of the large number of capabilities required for the production of more complex products.

Our model is flexible in that other policies can be simulated as well. Our choice for vertical policy as the addition of one new capability and horizontal policy as the unit improvement of maximum product complexity are ideal-types of vertical and horizontal policies, respectively. In between the two policies, one can put hybrid policies. Two such policies follow naturally from our model.

First, rather than viewing vertical policy as the addition of some random capability, one could further specify a vertical policy as one that specifically targets a capability that, following our model, can be recombined with already existing capabilities in ways that would maximize the increase in average product complexity in the economy. For low-income countries with few capabilities, the targeting of such capabilities may be relatively easy to gauge as the increase in the number of new recombinations resulting from one new capability, is still rather limited. For high-income countries with many capabilities, such a targeted vertical policy may be harder to determine. Yet, the underlying idea of targeting capabilities that can be recombined in many and complex ways clearly speaks to the logic of focusing on 'general purpose technologies' (as the term suggests) \citep{Bresnahan1995}.

Second, rather than viewing horizontal policy as a unit increase in the maximum product complexity that an economy can produce, one can imagine a more hybrid policy in which a government seeks to improve the maximum product complexity only in a certain broad sector (like healthcare, mobility, agriculture, etc.). In the model, sectors would correspond to a subset of products that would fall within sectoral boundaries. This would mean that horizontal policies can be made more specific to coordination challenges in certain sectoral contexts rather than across the board. Such policies remain horizontal in nature, but targeted in their scope. In particular, a policy maker would wish to target those sectors for which many relevant capabilities are already present, but which fail to leverage those capabilities in complex product due to present limits to coordination failures. This type of policy has also been discussed in the innovation policy literature under the heading of 'systemic policy' \citep{Smits2004, Wieczorek2012}. 

Finally, turning to the revival of industrial policy as a form of vertical policy, our model provides both support and a critique to industrial policy as a means to spur economic development. For low-income countries, there is a strong rationale for industrial policy as to increase their capability base. For such countries, focusing only on improving the ability to coordinate many capabilities makes little sense as long the number of capabilities present is still low. For high-income countries, however, the rationale for modern vertical policy is less obvious. As such countries can only compete on complex products with high value-added, the main challenge for these countries is to improve the ability to produce more complex products from the large set of capabilities that they already master. Here, horizontal policies alone could be, theoretically, sufficient to continue economic development. The exact distinction between policies for low-income countries and high-income countries could be determined empirically by looking at the inverted-U patterns between average income and product variety (with maximal variety located around 24,000 US Dollar (PPP in constant 2000) \citep{Cadot2013}). As countries go through this ‘hump’, they should start focusing more on horizontal policies.

In this light, the plea for industrial policy in the context of high-income countries, and equally for technology missions \citep{Mazzucato2011, Mazzucato2015}, needs more grounding. If such missions are articulated in terms of the alleged need to master a specific new technology domain or industry, our model would cast doubt about its effects on growth. While a new technological capability could indeed be beneficial for growth, it will generate little comparative advantage if actors within the innovation system are not able to combine and integrate the new capability with the existing set of capabilities, including complementary technologies, skills and institutions.

\section*{Acknowledgments}
Both authors are funded by the Netherlands Organisation for Scientific Research (NWO) under the Vici scheme, number 453-14-014.


\bibliographystyle{model2-names}
\bibliography{library.bib}

\end{document}